\documentclass[showpacs,twocolumn,aps,prl]{revtex4}
\usepackage{amsmath}
\usepackage{amsfonts}
\usepackage{amssymb}
\usepackage{amsthm}
\usepackage{graphicx}

\begin{document}

%----------------------------------------------------------------%
\title{Fast domain wall propagation under an optimal field pulse in magnetic
nanowires}
\author{Z. Z. Sun}
%\email[To whom correspondence should be addressed. Electronic
%address: ]{zhouzhou.sun@physik.uni-regensburg.de}
\affiliation{Institute for Theoretical Physics, University of Regensburg,
D-93040 Regensburg, Germany}
\author{J. Schliemann}
\affiliation{Institute for Theoretical Physics, University of Regensburg,
D-93040 Regensburg, Germany}
\date{\today}

\begin{abstract}
We investigate field-driven domain wall (DW) propagation in magnetic nanowires
in the framework of the Landau-Lifshitz-Gilbert equation.
We propose a new strategy to speed up the DW motion in a uniaxial magnetic
nanowire by using an optimal space-dependent field pulse synchronized
with the DW propagation. Depending on the damping parameter,
the DW velocity can be increased by about two orders
of magnitude compared to the standard case of a static uniform field.
Moreover, under the optimal field pulse, the change in total
magnetic energy in the
nanowire is proportional to the DW velocity,
implying that rapid energy release is essential
for fast DW propagation.
\end{abstract}
%\keywords{superlattice, SSCO, phase diagram}
\pacs{75.60.Jk, 75.75.-c, 85.70.Ay}
% 75.60.Jk Magnetization reversal mechanisms\\
%75.75.+a Magnetic properties of nanostructures \\
%85.70.Ay Magnetic device characterization, design and modeling \\
\maketitle
%----------------------------------------------------------------%
Recently the study of domain wall (DW) motion in magnetic nanowires has
attracted a great deal of attention, inspired both
by fundamental interest in nanomagnetism as well as
potential industrial applications.
Many interesting applications like memory bits\cite{Cowburn,Parkin} or
magnetic logic devices\cite{Allwood}
involve fast manipulation of DW structures, i.e. a
large magnetization reversal speed.

In general, the motion of a DW can be driven by a magnetic field
\cite{Ono, Atkinson,Beach}
and/or a spin-polarized current\cite{Klaui,Hayashi,Zhang,Tatara,Thiaville}.
Although the DW dynamics in systems of higher spatial dimension can be very
complicated, some simple but important results were obtained by
Schryer and Walker for effectively
one-dimensional (1D) situations\cite{Walker}:
At low field (or current density), the DW velocity $v$ is linear in the
field strength $H$ until $H$ reaches a so-called Walker breakdown
field $H_w$\cite{Walker}. Within this linear regime,
DW propagates as a rigid object.
For $H>H_w$, the DW loses its rigidity and develops a complex time-dependent
internal structure. The velocity can even oscillate with time due to the
``breathing'' of the DW width. The time-averaged velocity $\bar{v}$ decreases
with the increase of $H$, resulting in a negative differential mobility.
$\bar{v}$ can be again linear with $H$ approximately when $H \gg H_w$.
The predicted $v$-$H$ characteristic is in a good agreement with
experimental results on permalloy nanowires\cite{Ono,Atkinson,Beach}.
Recently a general definition of the DW velocity proper for any types of DW
dynamics has been also introduced\cite{Wang}.

For a single-domain magnetic nanoparticle (called Stoner
particle), an appropriate time-dependent but spatially homogeneous
field pulse can
substantially lower the switching field and increase the reversal speed since
it acts as an energy source enabling to overcome the energy barrier
for switching the spatially constant magnetization\cite{Sun, Wang1}. In the
present letter, we investigate the dynamics of a DW
in a magnetic nanowire under a field
pulse depending both on time and space. As a result, such a pulse,
synchronized with the DW propagation, can dramatically increase the DW
velocity by typically two orders compared with the situation of a constant
field. Moreover, the total magnetic energy typically decreases with a rate
being proportional to the DW velocity, i.e. the external field source can even
absorb energy from the nanowire.

\begin{figure}[htbp]
 \begin{center}
\includegraphics[width=7.5cm, height=3.cm]{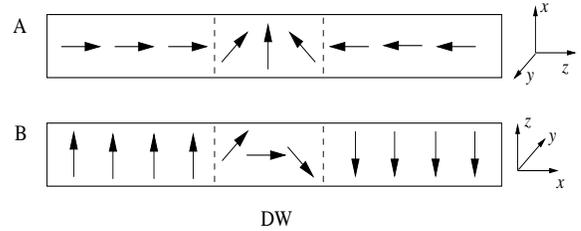}
 \end{center}
\caption{\label{fig1} A schematic diagram of two dynamically equivalent 1D magnetic
nanowire structures. (A) Easy axis is along the wire axis (z-axis); (B)
Easy axis (z-axis) $\perp$ the wire axis (x-axis). The region between two
dashed lines denotes the DW region. }
\end{figure}

%----------------------------------------------------------------%
A magnetic nanowire can be described as an effectively 1D continuum of magnetic
moments along the wire axis direction. Magnetic domains are formed due
to the competition between the anisotropic magnetic energy and the exchange
interaction among adjacent magnetic moments. Let us first concentrate
on the case of a uniaxial magnetic anisotropy:
Two dynamically equivalent configurations of 1D uniaxial
magnetic nanowires are schematically shown in Fig.~\ref{fig1}.
Type A shows the wire axis to be also the easy-axis (z-axis). Type B shows the
easy axis (z-axis) is perpendicular to the wire axis (x-axis).
Although our results described below apply to both configurations,
we will focus in the following on type B. The
spatio-temporal dynamics of the magnetization
density $\vec{M}(x,t)$ is governed by the
Landau-Lifshitz-Gilbert (LLG) equation\cite{llg}
\begin{equation}
\frac{\partial \vec{M}}{\partial t}= -|\gamma|\vec{M}\times\vec{H}_{t}
+\frac{\alpha}{M_s} \left (\vec{M}\times \frac{\partial \vec{M}}{\partial t}\right ),\label{LLG}
\end{equation}
where $|\gamma|=2.21 \times 10^5 (rad/s)/(A/m)$ the gyromagnetic ratio,
$\alpha$ the Gilbert damping coefficient, and  $M_s$ is the saturation
magnetization density. The total effective field $\vec{H}_{t}$ is given by
the variational derivative of the total energy with respect to magnetization,
$\vec{H}_{t}=-(\delta E/ \delta \vec{M})/ \mu_0$, where $\mu_0$
the vacuum permeability. The total energy
$E=\int_{-\infty}^{\infty}dx\varepsilon(x)$
can be written as an integral over an energy density (per unit section-area),
\begin{equation}
\varepsilon(x)=
-KM_z^2 +J\left [\left (\frac{\partial \theta}{\partial x}\right )^2
+\sin^2\theta \left (\frac{\partial \phi}{\partial x}\right )^2\right ]
- \mu_0 \vec{M} \cdot \vec{H},\label{energy}
\end{equation}
where $x$ is the spatial variable in the wire direction. Here $K$, $J$
are the coefficients of energetic anisotropy and exchange interaction,
respectively, and $\vec{H}$ is the external magnetic field.
Moreover, we have adopted the usual spherical coordinates,
$\vec{M}(x,t)=M_s(\sin\theta\cos\phi, \sin\theta\sin\phi, \cos\theta)$
where the  polar angle $\theta(x,t)$ and the azimuthal angle $\phi(x,t)$
depend on position and time.

Hence, the total field $\vec{H}_t$ consists of three parts: the external field
$\vec{H}$, the intrinsic uniaxial field along the easy axis
$\vec{H}^K= (2KM_z/\mu_0)\hat{z}$, and the exchange field
 $\vec{H}^J$ which reads in spherical coordinates as\cite{Walker, Hickey},
\begin{align}
H^J_{\theta} &= \frac{2J}{\mu_0M_s}\frac{\partial^2 \theta}{\partial x^2}
- \frac{J\sin 2\theta}{\mu_0M_s}
\left (\frac{\partial \phi}{\partial x}\right)^2,\nonumber\\
H^J_{\phi} &=
 \frac{2J}{\mu_0M_s\sin\theta}\frac{\partial }{\partial x}
\left(\sin^2 \theta \frac{\partial \phi}{\partial x}\right).
\end{align}
Following Ref.~\cite{Walker}, let us focus on DW structures fulfilling
$\partial \phi/\partial x =0$, i.e.
 all the magnetic moments rotate around the easy axis synchronously.
Then the dynamical equations take the form
\begin{align}
\Gamma&\dot{\theta} = \alpha \left(H_{\theta} -\frac{KM_s}{\mu_0}\sin 2\theta
+\frac{2J}{\mu_0M_s}\frac{\partial^2 \theta}{\partial x^2}\right) +H_{\phi},
\nonumber\\
\Gamma&\sin\theta\dot{\phi} = \alpha  H_{\phi}-H_{\theta}
+\frac{KM_s}{\mu_0}\sin 2\theta-\frac{2J}{\mu_0M_s}\frac{\partial^2 \theta}{\partial x^2}, \label{llg1}
\end{align}
where we have defined $\Gamma \equiv (1+\alpha^2)|\gamma|^{-1}$, and $H_i (i=r,\theta,\phi)$ are the three components of the external field in
spherical coordinates.
In the absence of an external field, an exact solution for a static DW
is given by $\tan {\frac{\theta(x)}{2}} =\exp(x/\Delta)$
where $\Delta=\sqrt{J/(KM_s^2)}$ is the width of the DW.
We note that a static DW can exist in a constant field only if
the field component along the easy axis is zero,  $H_z=0$.
In fact, according to Eqs.~\eqref{llg1} static solutions need to fulfill
 $H_{\phi}=0$ [implying $\phi=\tan^{-1}(H_y/H_x)$ is spatially constant] and
\begin{equation}
\frac{2J}{\mu_0M_s}\frac{\partial^2 \theta}{\partial x^2}
-\frac{KM_s}{\mu_0}\sin 2\theta+H_{\theta}=0
\end{equation}
or, upon integration,
\begin{equation}
\frac{J}{\mu_0M_s}\left(\frac{\partial \theta}{\partial x}\right)^2+\frac{KM_s}{2\mu_0}\cos 2\theta+H_r(\theta)={\rm constant}.
\end{equation}
Considering the two boundaries at $\theta=0 (x\rightarrow-\infty)$ and
$\theta=\pi (x\rightarrow +\infty)$ for the DW, we conclude $H_r(0)=H_r(\pi)$,
which requires $H_z =0$. In this case, the stationary DW solutions under a
transverse field are described
as $x=\int [\sqrt{(KM_s^2 \sin^2\theta- \mu_0M_sH \sin\theta)/J} ]^{-1}d \theta$.

%-----------------------------------------------------------------%
Thus, when an external field with a component along the easy axis is applied to the nanowire, the DW is expected to move. We use a travelling-wave
{\it ansatz} to describe rigid DW motion\cite{Walker},
\begin{equation}
\tan \frac{\theta(x,t)}{2} =\exp\left(\frac{x-vt}{\Delta}\right),
\end{equation}
where the DW velocity $v$ is assumed to be constant.
Substituting this trial function into Eq.~\eqref{llg1},
the dynamic equations become
\begin{equation}
\Gamma\sin\theta v = -\Delta (\alpha H_{\theta} +H_{\phi}),\quad
\Gamma\sin\theta\dot{\phi} = \alpha H_{\phi}-H_{\theta}.\label{main}
\end{equation}
Eq.~\eqref{main} describes the dependence of the linear velocity $v$
and the angular velocity
$\dot{\phi}$ on the external field $\vec H$.
Our following results discussion will be based on Eqs.~\eqref{main}.

%----------------------------------------------------------------%
Let us first turn to the case of a static field case
applied along the easy axis (z-axis in type B of Fig.~\ref{fig1}),
$H_{\theta}=-H \sin\theta, H_{\phi}=0$. Here we recover the well-known
static solution for a uniaxial anisotropy\cite{Slonczewski},
\begin{equation}
v=\frac{|\gamma|\Delta H}{\alpha+\alpha^{-1}},
\end{equation}
where the
azimuthal angle $\phi (t) = \phi(0)+|\gamma| H t/(1+\alpha^2)$
is spatially constant (i.e. $\partial \phi/\partial x =0$) and increases linearly
with time.

\begin{figure}[htbp]
 \begin{center}
\includegraphics[width=6.5cm, height=4.5cm]{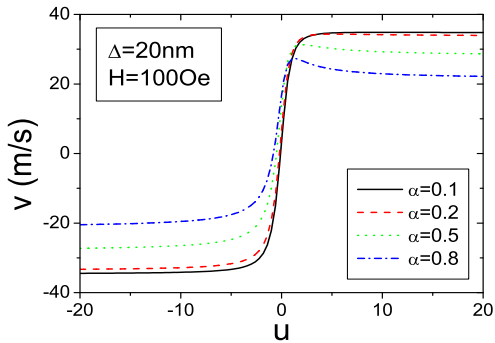}
 \end{center}
\caption{\label{fig2} (Color online) The DW propagation velocity $v$ versus the parameter $u$ at the different damping values $\alpha=0.1, 0.2, 0.5, 0.8$. The other parameters are chosen as $\Delta=20nm$ and $H=100{\rm Oe}$. }
\end{figure}

Let us now allow  the applied external field  to depend both on
space and time. Our task is to design, under a fixed field magnitude $H$, an
optimal field configuration $\vec{H}(x,t)$ to increase the DW velocity as
much as possible. From Eqs.~\eqref{main}, we find a manifold of
solutions of specific space-time field configurations described by a parameter
$u$,
\begin{align}
H_r(x,t)=&H\cos\theta,\quad H_{\theta}(x,t)=- H\sin\theta/\sqrt{1+u^2},\nonumber\\
&H_{\phi}(x,t)=-uH\sin\theta/\sqrt{1+u^2}.\label{pulse1}
\end{align}
%where $H_i (i=r,\theta,\phi)$ are the three components of the field pulse in
%spherical coordinates.
The velocities $v$ and $\dot{\phi}$ reads
\begin{equation}
v=\frac{|\gamma|\Delta H}{1+\alpha^{2}} \frac{\alpha+u}{\sqrt{1+u^2}},\quad
\dot{\phi} =\frac{|\gamma|H}{1+\alpha^{2}}\frac{1-\alpha u}{\sqrt{1+u^2}}.
\label{general}
\end{equation}
The previous static field case is recovered for $u=0$.
The maximum of the velocity $v_m$ with regard to $u$ us reached for
$u=1/\alpha$,
\begin{equation}
v_m=\frac{|\gamma| \Delta H}{\sqrt{1+\alpha^2}},
\end{equation}
where the angular velocity is zero, $\dot{\phi}=0$. On the other hand,
$\dot{\phi}$ attains a maximum for $u=-\alpha$, where, in turn, the
linear velocity vanishes.
In Fig.~\ref{fig2} we have plotted the dependence of the velocity on the
parameter $u$ for different damping strengths and typical values for the
DW width $\Delta$ and the magnitude $H$ of the external field.

%----------------------------------------------------------%
To understand the physical meaning of the maximum velocity $v_m$,
we note that, according to Eqs.~\eqref{main},  the field components
$H_{\theta}$ and $H_{\phi}$ are required to be proportional to
$\sin\theta$ to ensure the constant velocity under the rigid DW approximation.
Moreover, at $u=1/\alpha$ we have $H_{\theta}=\alpha H_{\phi}$, and from the
identity
\begin{equation}
(\alpha
H_{\theta}+H_{\phi})^2 +(\alpha H_{\phi}-H_{\theta})^2=(1+\alpha^2) (H^2-H_r^2),
\label{con}
\end{equation}
we conclude that the term $(\alpha H_{\theta}+H_{\phi})$ is maximal resulting in
a maximal velocity according to Eqs.~\eqref{main}.
As a result, the new velocity under the optimal field pulse
is larger by a factor of
$v_m/v=\sqrt{1+\alpha^2}/\alpha\approx 1/\alpha$ compared to a constant
field with the same field magnitude. To give a practical example, the typical
value for the damping parameter in permalloy is $\alpha =0.01$ which
results in an increase of the DW velocity by a factor of $100$.

\begin{figure}[htbp]
 \begin{center}
\includegraphics[width=6.5cm, height=4.5cm]{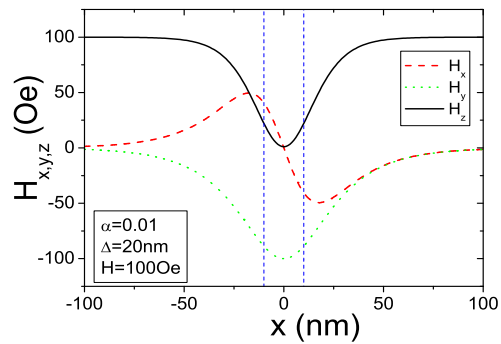}
 \end{center}
\caption{\label{fig3} (Color online) The $x, y, z$ components of the optimal
field pulse. The parameters are chosen as $\alpha=0.01$, $\Delta=20nm$ for
permalloy\cite{Beach}. The field magnitude is $H=100{\rm Oe}$. }
\end{figure}

It is instructive to also analyze the optimal field pulse
according to Eq.~\eqref{pulse1} with $u=1/\alpha$ in its cartesian
components,
\begin{align}
&H_x(x,t)= H\sin 2\theta (1-\alpha/\sqrt{1+\alpha^2})/2,\nonumber\\
&H_{y}(x,t)=- H\sin\theta/\sqrt{1+\alpha^2},\label{pulse}\\
&H_{z}(x,t)= H(\cos^2 \theta +\alpha \sin^2 \theta/\sqrt{1+\alpha^2}),\nonumber
\end{align}
where $\theta$ follows the wave-like motion $\tan \frac{\theta(x,t)}{2} =\exp(\frac{x}{\Delta}- \frac{ |\gamma| H}{\sqrt{1+\alpha^2}}t)$.
In Fig.~\ref{fig3} we plotted these quantities at $t=0$ around
the DW center where the main spatial variation of the pulse occurs.
Note that the space-dependent field distribution should move with the same
speed $v_m$ synchronized with the DW propagation.
Near the DW center the components $H_x$ and $H_z$ are (almost) zero whereas
a large transverse component $H_y$ is required to achieve fast DW propagation.
Qualitatively speaking, the transverse field causes a precession
of the magnetization resulting in its reversal. This finding is consistent
with recent micromagnetic simulations showing that
the DW velocity can be largely increased by applying an
additional transverse field\cite{transverse}.

It is also interesting to study the energy variation under the optimal field
pulse,
\begin{equation}
\frac{dE}{dt}=-\mu_0\int_{-\infty}^{+\infty} dx\left (\frac{\partial \vec{M}}{\partial t} \cdot \vec{H}_{t} +\vec{M} \cdot \frac{\partial \vec{H}}{\partial t}\right )\equiv P_{\alpha} + P_{h}.
\end{equation}
The first term $P_{\alpha}$ is the intrinsic damping power due to all kinds of
damping mechanisms described by the phenomenological parameter $\alpha$.
According to the LLG equation
$P_{\alpha}=-\frac{\mu_0\alpha}{|\gamma|M_s}
\int_{-\infty}^{+\infty} dx (\frac{\partial M}{\partial t})^2$ is always negative\cite{Sun},
implying an energy loss.
$P_{h}$ is the external power due to the time-dependent external field.
From Eq.~\eqref{general}, both powers are obtained as
\begin{align}
&P_{\alpha} = -\frac{2\alpha}{1+\alpha^2}\mu_0 |\gamma| M_s \Delta H^2,\label{pa}\\
P_{h} = &\frac{2\alpha(\sqrt{1+u^2}-1)-2u}{(1+\alpha^2)\sqrt{1+u^2}}\mu_0 |\gamma| M_s \Delta H^2, \label{ph}
\end{align}
such that the total energy change rate is
\begin{equation}
\frac{dE}{dt} = -2\mu_0 M_s H v=- \frac{ 2(\alpha+u)\mu_0 |\gamma| M_s \Delta H^2}{(1+\alpha^{2})\sqrt{1+u^2}}. \label{dedt}
\end{equation}
Note that the intrinsic damping power is independent of the
parameter $u$ and always negative, whereas the total energy change rate
is proportional to the negative DW velocity. Thus, for positive
velocities ($u>-\alpha$) the total
magnetic energy decreases while it grows for negative velocities ($u<-\alpha$).
In the former case energy is absorbed by the external field source while
in the latter case the field source provides energy to the system.
The optimal field source  helps to rapidly release or gain magnetic energy
which is essential for fast DW motion.
This aspect is very different from the reversal of a Stoner particle where the
time-dependent field is always needed to provide energy to the system
to overcome the energy barrier\cite{Sun}.

%----------------------------------------------------------------%
Moreover, our new strategy of employing space-dependent field pulses
can also be applied to uniaxial
anisotropies of arbitrary type: Let $w(\theta)$ be
the uniaxial magnetic energy density. The static DW solution in the absence
of an external field reads
$x =\int \chi^{-1}(\theta) d\theta$, where
\begin{equation}
\chi(\theta) =\sqrt{[w(\theta)-w_0]/J}.
\end{equation}
Here $w_0$ is the minimum energy density for magnetization along the easy axis.
By performing analogous steps as before, we obtain the the optimal velocity as
$v_m=\frac{|\gamma| H}{\sqrt{1+\alpha^2}\chi_{max}}$, where $\chi_{max}$
denotes the maximum of $\chi(\theta)$ throughout all $\theta$.

On the other hand, our approach is not straightforwardly extended to
the case of a magnetic wire with biaxial anisotropy. To see this, consider,
a biaxial anisotropy $\varepsilon_i=-KM_z^2+K'M_x^2$
where the coefficients $K$, $K'$ correspond to the easy and hard axis,
respectively\cite{Walker}. The LLG equations read
\begin{align}
& \Gamma \dot{\theta} = \alpha \left ( H_{\theta} -\frac{KM_s}{\mu_0}\sin 2\theta -\frac{K'M_s}{\mu_0}\sin 2\theta\cos^2\phi \right.\nonumber\\
&\left.+\frac{2J}{\mu_0M_s}\frac{\partial^2 \theta}{\partial x^2} \right) +H_{\phi}+\frac{K'M_s}{\mu_0}\sin \theta\sin 2\phi,\nonumber\\
& \Gamma \sin\theta\dot{\phi} =\alpha H_{\phi}-H_{\theta}+\frac{KM_s}{\mu_0}\sin 2\theta -\frac{2J}{\mu_0M_s}\frac{\partial^2 \theta}{\partial x^2}\nonumber\\
& +\frac{K'M_s}{\mu_0}\sin 2\theta\cos^2\phi+ \frac{\alpha K'M_s}{\mu_0}\sin \theta\sin 2\phi.\label{bi}
\end{align}
Let us assume $\phi(x,t)=\phi_0$ is a constant determined by the applied
field. Substituting the travelling-wave {\it ansatz}
$\tan \frac{\theta(x,t)}{2} =\exp\left(\frac{x-vt}{\Delta}\right)$, where now
$\Delta=\sqrt{J/(K +K' \cos^2\phi_0)}/M_s$, into Eqs.~\eqref{bi} we obtain
\begin{align}
& \Gamma \sin\theta v= -\Delta(\alpha H_{\theta} +H_{\phi} +K'M_s\sin \theta\sin 2\phi_0/\mu_0),\label{bi1}\\
&  \alpha K' M_s\sin \theta\sin 2\phi_0/\mu_0 + (\alpha  H_{\phi}-H_{\theta})=0.\label{bi2}
\end{align}
For a static field along z-axis $H_{\theta}=-H \sin\theta, H_{\phi}=0$, the
solution is just the Walker's result $v=|\gamma|\Delta H/\alpha$
(Note here $\Delta$ also depends on $H$)\cite{Walker}. To implement our
new strategy, we need to find the maximum of the right-hand side of
Eq.~\eqref{bi1} under two constraints of Eq.~\eqref{bi2} and
Eq.~\eqref{con} with $H_{\theta}$ and $H_{\phi}$ being proportional to $\sin\theta$.
The unique solution to this problem is indeed a constant field along the
z-axis which is thus the optimal field configuration.

%------------------------------------------------------%
In summary, our theory is general and can be applied to a magnetic nanowire with a uniaxial anisotropy which can be from shape, magneto-crystalline or the dipolar interaction. The experimental challenge of our proposal is obviously the generation
of a field pulse focused on the DW region and synchronized with its motion.
However, the field source synchronization velocity can be pre-calculated from the material parameters. As for the required localized field (See Fig.~\ref{fig3}), we propose to employ a ferromagnetic scanning tunneling microscope (STM) tip to produce a localized field perpendicular to the wire axis\cite{Michlmayr} and use a localized current to produce an Oersted field along the wire axis\cite{Michlmayr1}. Moreover, such required localized fields may also be produced by
nano-ferromagnets with strong ferromagnetic (or antiferromagnetic)
coupling to the nanowire.
%However, in order to substantially speed up  the DW motion
%one does not necessarily need to fulfil the optimum field configuration
%since the DW velocity, as seen in Fig.~\ref{fig2} is a rather flat function
%of the parameter $u$ for $u\gtrsim 1/\alpha$.
%Thus, also
%other time-dependent field configurations close to the one-parametric
%manifold given in Eq.~(\ref{pulse1}) will lead to a substantial
%increase of the DW velocity.
We also point out that, although the field source typically does not
consume energy but gain energy from the magnetic nanowire, the pulse source
may still require excess energy to overcome effects such as defects pinning,
which is not included in our model.
At last, the generalization of the strategy beyond the rigid DW
approximation, and to DW motion induced
by spin-polarized current
will also be attractive direction of future research.
%In conclusion, we have proposed a new strategy to speed up DW propagation
%in uniaxial magnetic nanowires. Our approach employs a space-dependent
%field pulse synchronized with the DW motion.
%The DW velocity can be increased
%dramatically, depending on the damping parameter, by two orders of
%magnitude compared with the standard case of a static uniform field.

Z.Z.S. thanks the Alexander von
Humboldt Foundation (Germany) for a grant.
This work has been supported by Deutsche Forschugsgemeinschaft via SFB 689.

\end{document}